# Cavity optomechanics on a microfluidic resonator with water and viscous liquids

**Kyu Hyun Kim[1+], Gaurav Bahl[2+], Wonsuk Lee[1,3], Jing Liu[3], Matthew Tomes[1], Xudong Fan[3], and Tal Carmon[1*].**

[1]*Department of Electrical Engineering and Computer Science, University of Michigan*

*1301 Beal Ave. Ann Arbor, MI, 48109, USA*

[2]*Mechanical Science and Engineering, University of Illinois at Urbana-Champaign,*

*1206 W. Green St., Urbana, IL 60801*

[3]*Department of Biomedical Engineering, University of Michigan*

*1101 Beal Ave. Ann Arbor, MI, 48109,USA*

*e-mail: tcarmon@umich.edu

+Equally contributing authors

**Currently, optical- *or* mechanical-resonances are commonly used in microfluidic research. However, optomechanical oscillations by light pressure were not shown with liquids. This is because replacing the surrounding air with water inherently increases the acoustical impedance and hence the associated acoustical radiation-losses. Here, we bridge between microfluidics and optomechanics by fabricating *hollow* bubble resonators with liquid *inside* and optically exciting 100-MHz vibrations with only mW optical-input power. This constitutes the first time that any microfluidic system is optomechanically actuated. We further prove the feasibility of microfluidic optomechanics on liquids by demonstrating vibrations on organic fluids with viscous-dissipation higher than blood viscosity while measuring density changes in the liquid via the vibration frequency shift. Our device will enable using cavity optomechanics for studying non-solid phases of matter.**

A major application of *optical* resonators is in the field of sensing where microresonators were used to sense biomarkers in serum (*1*) and to detect viruses and nanoparticles (*2-4*) in an aqueous environment. With similar motivation and in a parallel effort, *mechanical* resonators were used to weigh biomolecules and cells *(5-7)*. Just like we use more than one sense (e.g. eyes and ears) to detect hazards, optomechanics might suggest a bridge between the seemingly parallel optical- and mechanical-detection fields. Optomechanics allows excitation and interrogation of GHz rate vibrations in a well confined region in space while allowing measurement via a single optical output. These GHz acoustical bands were rarely investigated in liquids. The recent availability of liquid containing bubble shaped resonators *(8, 9)*, in combination with optomechanical vibrations at >GHz rate (*10-12*) might pave the way for ultrasound investigations on analytes in liquids using micron-scale acoustical wavelengths. Nevertheless, cavity optomechanics on non-solid phases of material was never before demonstrated.

One of the major "show stoppers" on the way to microfluidic optomechanics originates from the fact that water has acoustical impedance that is more than 4000 times larger than air. Hence, naively immersing optomechanical devices in water will accordingly increase the acoustical radiation losses. Liquid submerged optomechanical oscillators are therefore challenging, as sound will tend to escape from the cavity by radiating out rather than being confined to the resonator. Here we make a transformative change by confining the high-impedance water *inside (6, 13)* a silica-bubble resonator so that its acoustical quality factor is minimally affected. In this manner, when both mechanical and optical quality factors, $Q_m$ and $Q_O$, of microdevices are leveraged to be high enough, their optical and mechanical modes can be parametrically coupled allowing the optical excitation of vibrations (*14*). In spherical shapes (*10*) such as our bubble, excitation of mechanical radial breathe modes is expected via the centrifugal pressure applied by a circulating optical whispering-gallery mode. The optical pressure here is related to the fact that photons carry linear momentum, resulting in a force similar to what we feel when our car makes a sharp curve. It is interesting that we could also observe some 1 GHz and 11 GHz modes in the same microfluidic device that were probably excited via a different mechanism as explained in (*12, 15*)

Bubbles along micro-capillaries (*8, 9, 16, 17*) and on-chip (*18*) devices can possess excellent optical transmission (*8, 9, 16, 17*) together with low mechanical dissipation while allowing liquid to be placed inside the device. Additionally, such glass bubbles can benefit from optically excited vibrations as no electrodes are needed. This is because optical excitation can be performed in a contact-less manner with light coupled into the device from a nearby tapered fiber (*19*). This makes such silica-glass bubbles attractive as a new type of a microfluidic optomechanical device as we will show here.

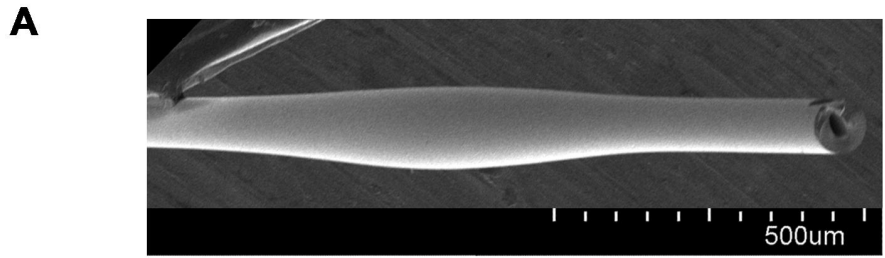

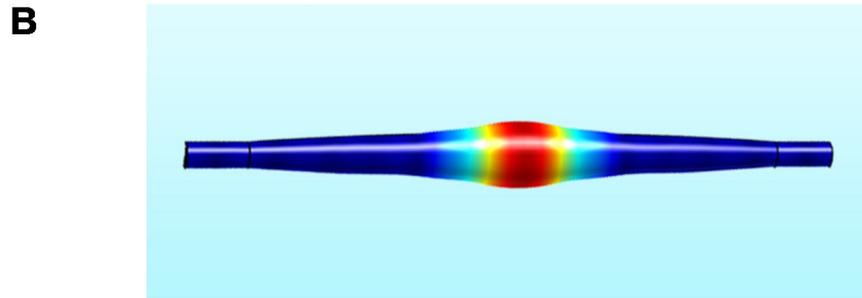

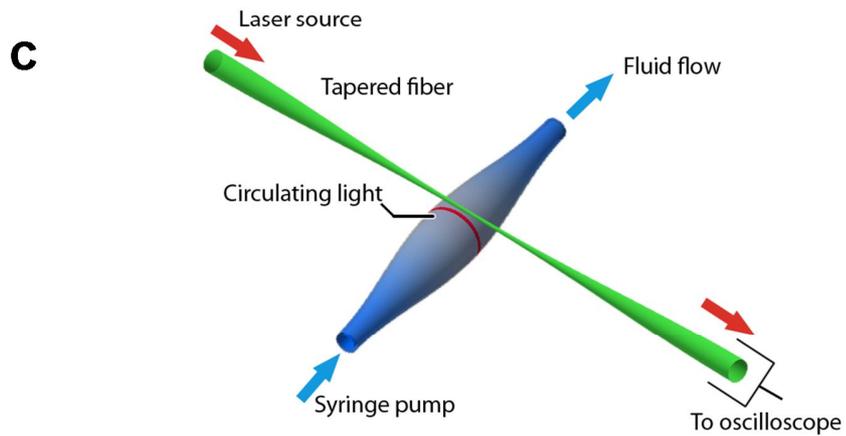

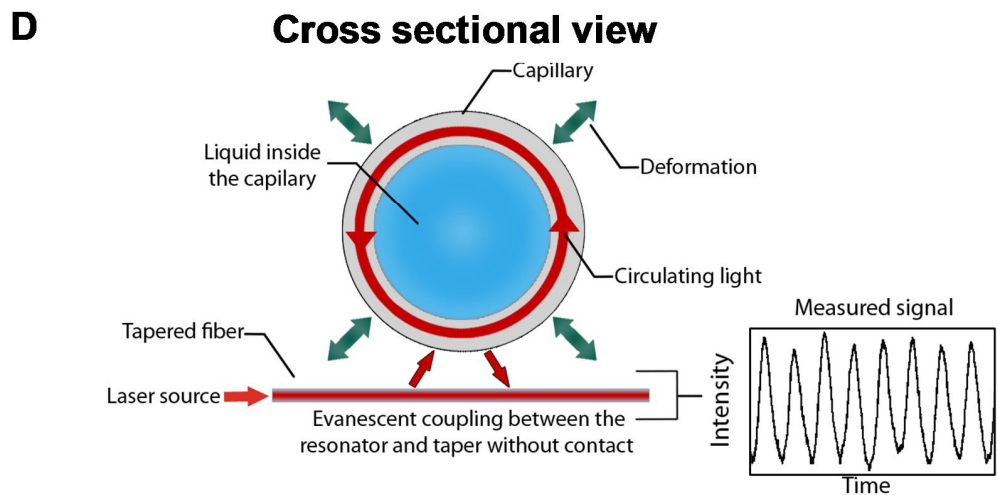

**Fig. 1. Fabrication and experimental setup** (A) SEM image of bubble-shaped capillary resonators together with its calculated eigen mechancial mode (B) where colors stands for deformation. Diameter and wall-thickness of our device are 72 and 10 μm. (C, D) Experimental setup: light (red) is evanescently coupled to an optical whispering-gallery mode of a bubble capillary where its centrifugal radiation pressure excites the bubble's mechanical breathe mode (D, green arrows). This vibration modulates the transmitted light as detected with a photodiode on the other side of the coupler. Transmitted signal shows high modulation due to mechanical vibrations of the resonator. Water is inserted in the bubble through a micro-capillary inlet, using a syringe pump. The optical wavelength that we used is 1.5 μm.

*Fabrication.-* Following the similar method presented in (8), we stretch a silica capillary while softening the glass using amplitude modulated $CO_2$ lasers, into a large diameter bubble in the middle of a micro-capillary (Fig. 1A, B). We can fabricate devices with various bubble diameters upon request and with wall thickness below 600 nm (*9*).

*Experimental setup.-* As shown in Fig. 1C and Fig. 1D, we evanescently couple light from a tapered optical fiber (*19*) into the glass bubble to excite its optical whispering-gallery mode. The inflating radiation pressure on the device walls is enhanced by the number of optical recirculations (finesse) that can reach a million, and excites the vibrational breathe mode of the bubble in a self sustained process (*14*). Vibration, in return, modulates the circulating light in a manner that can be observed by a photodetector. We optically measure mechanical vibrations by coupling part of the resonator light that was circulating in the vibrating cavity through the other side of fiber to a photodetector (Fig. 1D).

Light experiences a high degree of modulation (Fig. 1D) due to the mechanical vibrations and its power is at the mW scale, making it easily measurable with a low-noise photodetector and an electrical spectrum analyzer (Fig. 1C, Fig. 2). To reiterate, the mechanical mode is both excited and interrogated by light via standard telecom-compatible optical fibers with no external-feedback, control or modulation needed. As for the fluidic capability, liquids or gases are flown through the capillary by using a syringe pump.

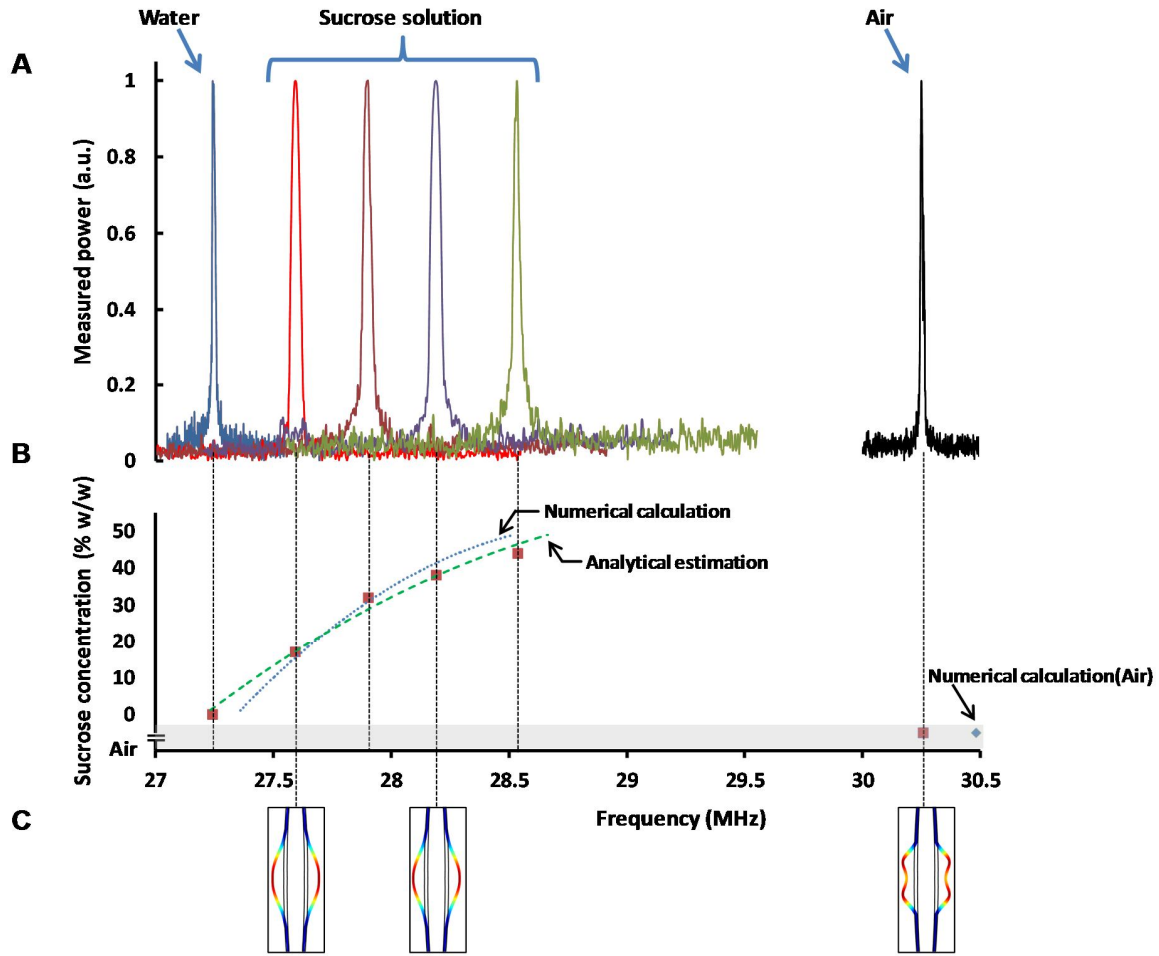

**Fig. 2. Proof of feasibility: optomechanical vibration on non-solid phases of materials** (A) Experimentally measured opto-excited vibration while different fluids are inside the resonator. The inspected fluids are air and viscous water-sugar solution with sugar concentration changing from 0 to 45% w/w. (B) The experimental frequency drift is presented as a function of the fluid density together with its corresponding finite-element numerical calculation and analytical estimation (Eq. 1). (C) finite element calculation of the mechanical modes where wireframe represents the device at mechanical equilibrium. Colors and deformation of the wall represent the acoustical amplitude. Note that despite of its shape; the "with air" mode is the first order mode as all of its sections are moving in phase.

*Experimental results.-* We start our experiment with a sugar-water solution at 45% weight concentration, and measure a 28.6 MHz vibration (Fig. 2A, green ). We choose to use high concentration aqueous-sucrose solution because its viscosity is 10 times higher than water and 3 times higher than blood, showing that our microfluidic optomechanical device [μFOM] works even on liquids with relatively large mechanical dissipation. To prove the feasibility of monitoring changes in the inspected analyte, we measured the vibration frequency while reducing the concentration of sugar down to pure water and then replacing water with air. A frequency shift of more than 3 MHz was observed. This large drift is not surprising since the fluid occupies a large portion of our device. We will now theoretically explain this frequency drift. First, we fit the measured frequency drift (Fig. 2A) to a simplified analytical model where the vibrational frequency, $f$, is related to the time it takes sound to cross from one side of the device to the other as given by

$$\frac{1}{f} = \frac{x_{sio2}}{v_{sio2}} + \frac{x_{sw}}{v_{sw}} \qquad (1)$$

where $v_{sio2}$ is the speed of sound in silica and $v_{sw}$ is the speed of sound in the water-sugar solutions as measured in (20). Similarly, $x_{sio2}$ and $x_{sw}$ represents the thickness of silica and liquid region. Optimizing $x_{sio2}$ and $x_{sw}$ as free parameters resulted a fit that is shown in the figure 2B in green, next to the corresponding experimental measurements. While Eq. 1 stands on a physical ground and follows the experimental results, a more complete model would provide additional details, including on the dynamics of silica and liquid area. We therefore calculated a numerical finite element solution that is simultaneously involving different physical models for the solid and liquid region of our device as explained in reference (21) ; such model is generally referred to as a multiphysics model. Our model provides the natural mechanical frequency and the corresponding deformation of the device as presented in figure 2. It is evident that adding liquid dramatically modifies the shape of the mechanical mode; again, this behavior is expected since liquid occupies most of the volume of our device and has density much higher than air.

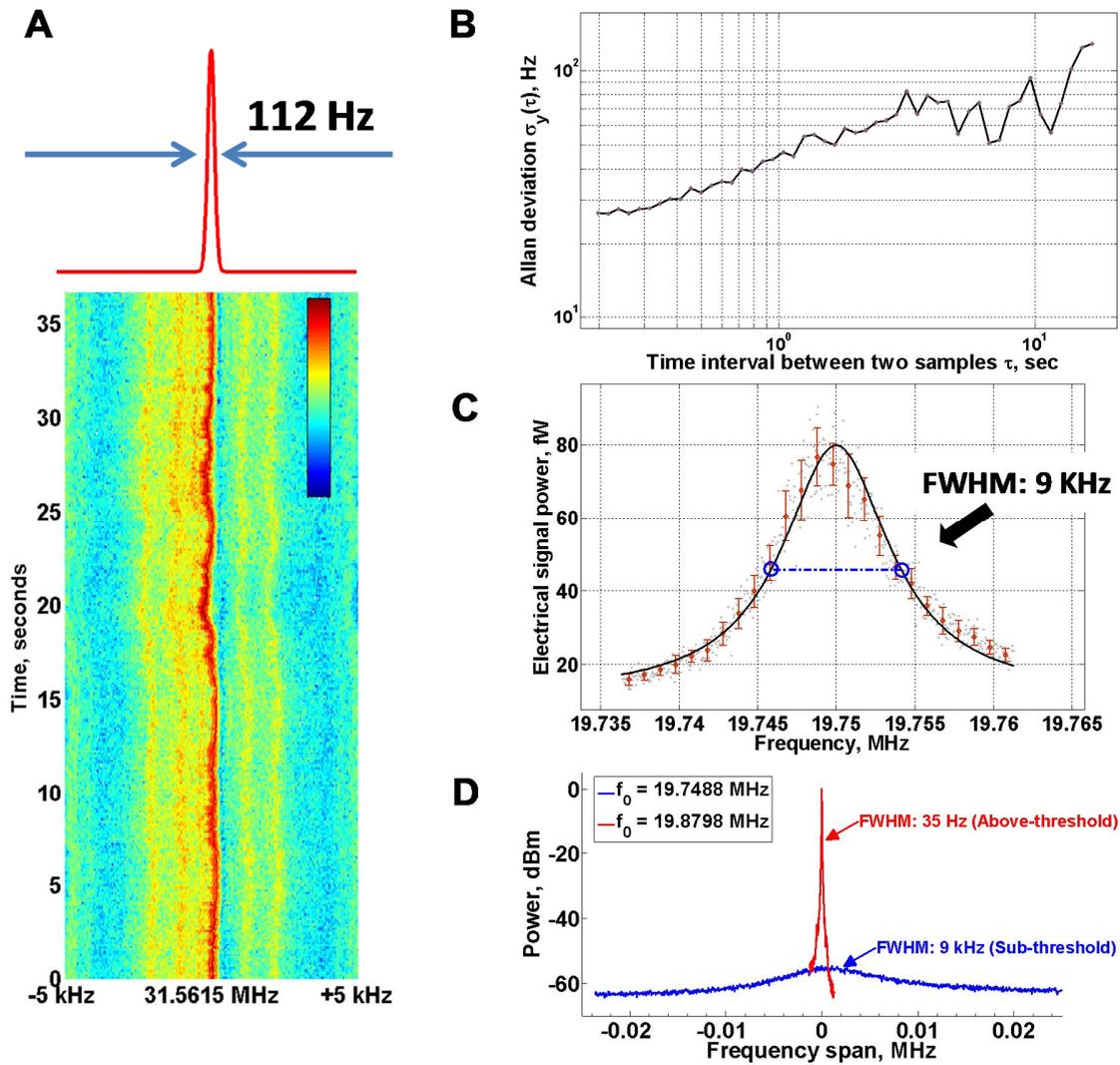

**Fig. 3.** (A) Measuring frequency stability of a water-filled resonator using a spectrogram. Standard deviation was measured as 112 Hz. (B) is Allan deviation of oscillation peak frequencies. (C) Measuring mechanical quality factor of a water-filled resonator was obtained from the resonance linewidth at sub-threshold. (D) In above-threshold behaviour, linewidth narrowing down to 35 Hz was measured.

Showing above that we can experimentally work with viscous liquid while measuring changes, we now go to the important question about the detection resolution.

Practically, the detection resolution is affected by the oscillation linewidth (such as the ones in Fig. 2) and by the frequency drift versus time. The oscillation full width at half max (FWHM) for a water filled bubble is measured to be 9 kHz near oscillation threshold (at 1 mW of optical input power ) implying a "cold" mechanical quality factor of 2172 at 19.75 MHz. When the optical input is increased over the oscillation threshold, the linewidth narrows to 35 Hz as shown in figure. The frequency drift of the oscillation is then measured in the form of a spectrogram (Fig. 3A). Statistics of the oscillation spectrogram over a 36 second period provide a standard deviation of 112 Hz (Fig. 3A). A more complete analysis for frequency stability is given by the Allan deviation *(22, 23)* (Fig. 3B), which calculates deviation between two frequency measurements separated by a fixed time difference. Being conservative, we estimate our resolution limit as the 112 Hz standard deviation rather than the smaller Allan deviation for intervals less than 10 second. We believe that in the future a differential detection scheme could cancel out drifts using a split mechanical mode similar to what is currently done in optical detection (4).

The optical quality factors, $Q_O$, are measured using the resonance linewidth to be near $7 \times 10^6$. This value of $Q_O$ is not affected by introducing water inside the bubble resonator. Since water absorbs light for the wavelength used (1.5 μm) this result indicates that the optical mode did not extend to the liquid. We can therefore conclude that the mechanical mode can be optically excited even with optically opaque liquids inside the capillary.

The mechanical quality factor, $Q_m$, with water inside, is measured from the sub-threshold mechanical linewidth. We measure $Q_m$ = 2,172 at *f* = 19.75 MHz (Fig. 3C) representing a *Q-f* coefficient of $4.3 \times 10^{10}$ Hz. This quality factor is six times smaller than what the dry device allows.

Measurement of the vibration threshold shows that oscillations typically start near 1 mW of optical power for the dry and wet capillaries. Oscillation is sustained as long as we keep our laser source on.

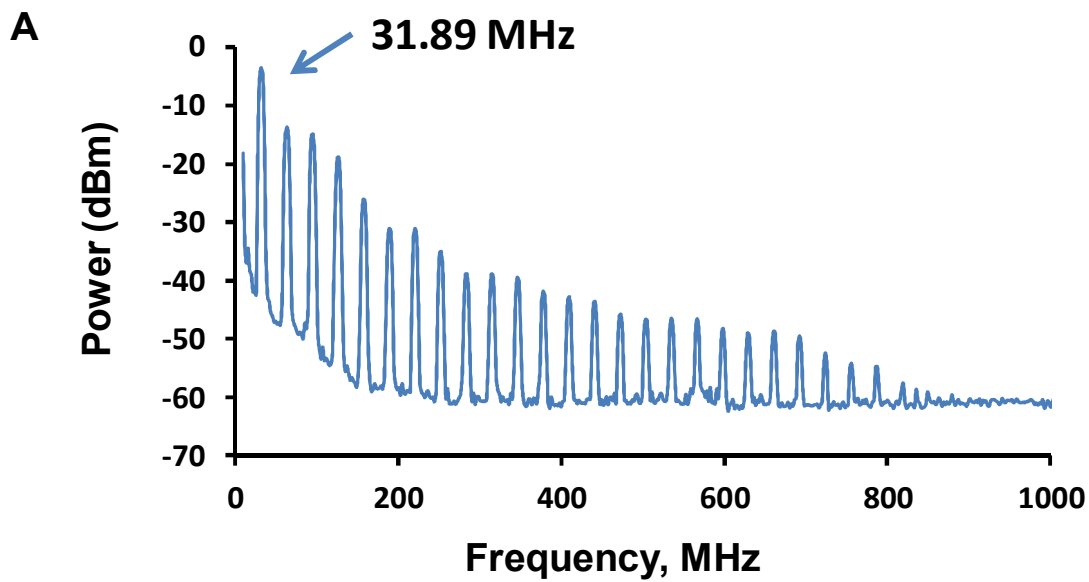

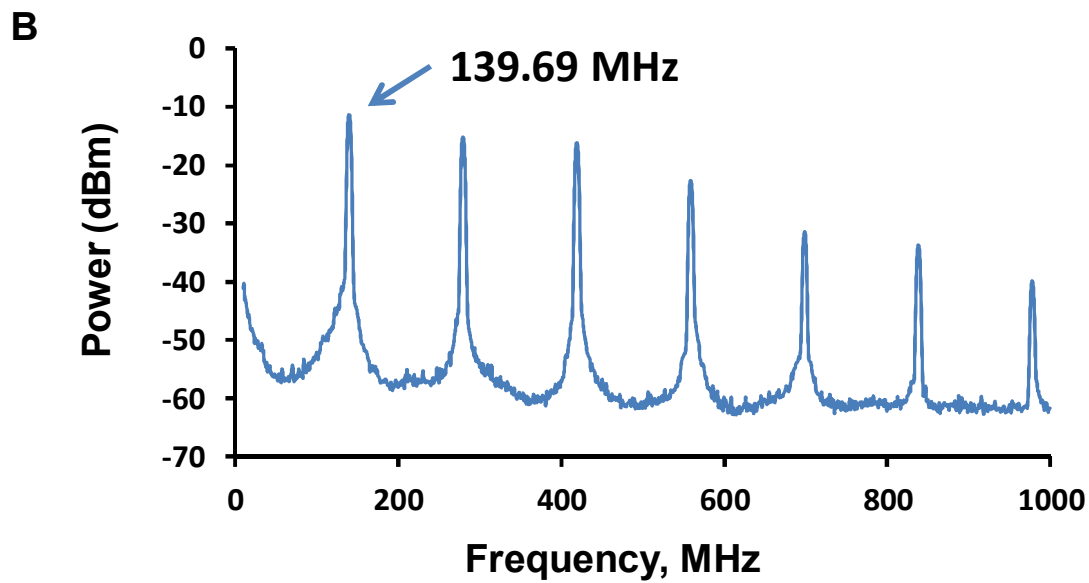

**Fig. 4. Experimental results in a water filled resonator:** (A, B) high-frequency modes are excited with many harmonics in the measured signal.

As the optical mode is phase modulated by the mechanical vibration, higher order optical (Stokes) lines exist in the transmitted light (*10*). This is evident with more than 25 Harmonics of a 32 MHz vibration spanning up to 0.8 GHz as shown in Fig. 4A. These high harmonics indicate that the mechanical vibrations result an optical frequency shift that is much larger than the optical resonance width (*10*) and might be useful in cases where frequency multiplication is needed without adding any hardware. Moreover, we were able to measure vibrations up to 140 MHz rates (Fig. 4B) while relying on high order mechanical modes (*10*). The harmonics of this mode signal were above 1 GHz in frequency. Additionally, we measured mechanical modes higher than GHz that we believe originates from a different mechanism as described in *(12, 15)*.

While in many optofluidic devices the motion of fluids affects light (*24, 25*), here, in contrast, radiation pressure allows light to create mechanical vibration in a device that contains a fluid. We show here sustainable optically excited vibrations in a μFOM bubble at 20 MHz to 140 MHz rates with different types of liquids inside the resonator. A variety of conditions can be relatively easily maintained inside the bubble, including fluids more viscous than serum, while light is conveniently coupled from the outer dry side of the bubble.

References

1   H. Zhu, P. S. Dale, C.W. Caldwell, X. Fan, *Anal. Chem.* **81,** 9858-9865 (2009)
2   F. Vollmer, S. Arnold, *Nature Methods* **5**, 7, 591 (2008)
3   T. Lu, K. J. Vahala et al., *Proc. Natl. Acad. Sci.* **108,** 15, 5976 (2011)
4   L. He, S. K. Özdemir, J. Zhu, W. Kim, L. Yang, *Nature Nanotechnol.* **6,** 428-432 (2011)
5   B. Ilic, D. Czaplewski, M. Zalalutdinov, H. G. Craighead, P. Neuzil et al., *J. Vac. Sci. Technol. B* **19**, 2825 (2001)

6	T. P. Burg, S. R. Manalis, *Nature* **446,** 1066-1069 (2007)
7	K. Y. Fong, W. H. P. Pernice, Mo Li, H. X. Tang, *Appl. Phys. Lett.* **97**, 073112 (2010)
8	M. Sumetsky, Y. Dulashko, R. S. Windeler, *Opt. Lett.* **35**, 7, 898 (2010)
9	W. Lee , X. Fan et al., *Appl. Phys. Lett.* **99,** 091102 (2011)
10	T. Carmon, K. J. Vahala, *Phys. Rev. Lett.* **98,** 123901 (2007)
11	M. Eichenfield, R. Camacho, J. Chan, K. J. Vahala, O. Painter, *Nature Lett.* **459** (2009)
12	G. Bahl, J. Zehnpfennig, M. Tomes, T. Carmon, *Nature Comm.* **2,** 403 (2011)
13	R. A. Barton, H. G. Craighead et al., *Nano Lett.* **10,** 2058-2063 (2010)
14	T. Carmon, H. Rokhsari, L. Yang, T. J. Kippenberg, K. J. Vahala, *Phys. Rev. Lett.* **94,** 223902 (2005)
15	M. Tomes, T. Carmon, *Phys. Rev. Lett.* **102,** 113601 (2009)
16	A. Watkins, J. Ward, Y. Wu, S. N. Chormaic, *Opt. Lett.* **36**, 11, 2113 (2011)
17	S. Berneschi, S. Soria et al., *Opt. Lett.* **36,** 17, 3521-3523 (2011)
18	E. J. Eklund, A. M. Shkel, *JMEMS.* **16,** 2, 232-239 (2007)
19	M. Cai, O. Painter, K. J. Vahala, *Phys. Rev. Lett.* **85,** 1, 74-77 (2000)
20	M. A. Rao, S. S. H. Rizvi, A. K. Datta, *Engineering Properties of Foods* (Taylor & Francis, CRC press, ed. 3, 2005), pp. 582-583. [third edition]
21	A.E. Baroudi, F. Razafimahery, L. Rakotomanana-Ravelonarivo, *Int. Jour. of Eng. Sci.*, **51**, 1-13 (2012)
22	D.W. Allan, *Proc. IEEE* **54**, 2, 221-231 (1966)
23	D.B. Sullivan, D.W. Allan, D.A. Howe, F.L. Walls, *NIST Tech Note* **1337** (1990)
24	D. Psaltis, S. R. Quake, C. Yang, *Nature* **442,** 381-386 (2006)
25	A. M. Stolyarov, Y. Fink, *Nature Photon.* **6,** 229-233 (2012)